\documentclass[conference]{IEEEtran}
\usepackage{graphicx}
\usepackage{setspace}
\usepackage{amssymb}
\usepackage{amsfonts}
\usepackage{mathrsfs}
\usepackage{amssymb,amsmath}
\usepackage{epstopdf}
\usepackage{bm}
\usepackage{algorithm}
\usepackage{algorithmic}
\usepackage{amsmath,amssymb,epsfig,graphics,subfigure}
\usepackage{theorem}
\usepackage{array,color}
\usepackage[compress]{cite}
\usepackage{stfloats}
\usepackage{graphicx}
\usepackage{subfigure}
\usepackage{flushend}
\usepackage{array}
\usepackage{cases}
\usepackage{multirow}
\usepackage{chngpage}
\usepackage{longtable}
\usepackage{amsfonts}
\usepackage{mathrsfs}
\usepackage{amssymb,amsmath}
\usepackage{algorithm}
\usepackage{algorithmic}
\usepackage{amsmath,amssymb,epsfig,graphics,subfigure}
\usepackage{theorem}
\usepackage{array,color}
\usepackage{amssymb}
\newcommand{\tb}{\mathbf}

\newcommand{\rmsg}{\overrightarrow}
\newcommand{\lmsg}{\overleftarrow}
\begin{document}
\title{{Joint Radar-Communication-Based Bayesian Predictive Beamforming for Vehicular Networks}}
\author{\IEEEauthorblockN{Weijie Yuan$^{1}$, Fan Liu$^{2}$, Christos Masouros$^{2}$, Jinhong Yuan$^{1}$, and Derrick Wing Kwan Ng$^{1}$}
\IEEEauthorblockA{
$^{1}$School of Electrical Engineering and Telecommunications, University of New South Wales, NSW 2052, Australia\\
$^{2}$Department of Electronic and Electrical Engineering, University College London, London, WC1E 7JE, UK\\
Email: \{weijie.yuan, j.yuan, w.k.ng\}@unsw.edu.au, fan.liu@ucl.ac.uk, chris.masouros@ieee.org}
}
\maketitle
\begin{abstract}
In this paper, we develop a predictive beamforming scheme based on the dual-functional radar-communication (DFRC) technique, where the road-side units estimates the motion parameters of vehicles exploiting the echoes of the DFRC signals. Compared to the conventional feedback-based beam tracking approaches, the proposed method can reduce the signaling overhead and improve the tracking performance. A novel message passing algorithm is proposed, which yields a near optimal performance achieved by the maximum \emph{a posteriori} estimation. Simulation results have shown the effectiveness of the proposed DFRC based scheme.
\end{abstract}


\section{Introduction}
In the future vehicular networks, the communication capability is of great importance to support various heterogenous applications including autonomous vehicles and traffic management \cite{wong2017key,chen2020massive}. In addition,  radar-type technologies are envisioned as a promising candidate for detecting and tracking cars and road obstacles in real-time due to the time-varying nature of the network topology and the surrounding environments. Traditionally, radar and communication systems exploit separate spectral resources and thus rarely interfere with each other. However, the separated approach becomes challenging in the future due to the limited available spectrum resources and increasing demand by both communities. As a result, the dual-functional radar communication (DFRC) technique, which performs both radar sensing and communication functionalities with a single transmission has been attracting attention over the last few years\cite{saddik2007ultra}.

Early contributions of DFRC designs focused on sub-6G Hz frequency band and cannot support Gbps data rate as required by vehicular communication systems.
To further improve the transmission rate, the bandwidth available in the millimeter wave (mmWave) spectrum serves as a key enabler for DFRC systems \cite{ZhaoLouMMwAVE}, which also improves the range resolution for radar. Aiming for designing DFRC transceivers at the mmWave band, the authors of \cite{liu2020joint} developed a novel framework based on hybrid analog-digital beamforming techniques. However, \cite{liu2020joint} did not take the high-mobility environments into account and their results are thus not suitable for vehicular applications. 
In mmWave systems, pencil-like spatial beams can be generated by a transmitter focusing the radiation power on the intended directions, which compensates the high path-loss of the mmWave signals. To establish a reliable communication link, it is essential to align the transmit and receive beams between the vehicles and the associated road-side unit (RSU) \cite{haghighatshoar2016beam}. To cope with the high-mobility constraint, several works considered the extended Kalman Filtering (EKF)-based fast beam tracking problem from the communication perspective\cite{zhang2019codebook,shaham2019fast}. In particular, the RSU first sends a communication signal containing pilots to the vehicles. Then the vehicles decode the information and estimate the relative angles with respect to the RSU which are then feedback to the RSU for beam steering. To achieve highly accurate estimation result, the number of pilots for EKF beam tracking should be sufficiently large, leading to prohibitively high communication signaling overhead. For these reasons, we aim to develop a low-overhead DFRC-based scheme for tracking the beam direction as well as the motion parameters in vehicular networks.

In this paper, we propose a novel DFRC-based predictive beamforming scheme for vehicle-to-infrastructure (V2I) scenarios. The prediction of the relative angles between the RSU and vehicles can reduce the latency for beam alignment. Moreover, with the help of the radar system, feedbacks from the vehicles to the RSU can be avoided. Compared to some feedback-based schemes that exploits a limited number of pilots for beam tracking, e.g. \cite{shaham2019fast}, the proposed DFRC-based scheme utilizes the whole downlink block both as communication data symbols and sensing pilots, which not only reduces the signaling overhead but improves the estimation performance. To determine the beam direction as well as other motion parameters, we introduce a specifically tailored factor graph-based framework and propose a low complexity message passing algorithm with parametric message representations. Simulation results show that compared to the communication-only feedback scheme, the proposed algorithm achieves better tracking performance and higher achievable rate.

	\emph{Notations:} We use a boldface letter to denote a vector. The superscripts $(\cdot)^{-1}$ {and} $(\cdot)^{\rm H}$ denote {the} inverse and the Hermitian operations{, respectively}; $\mathcal{N}(x;{m}_{x},{\lambda}_{x})$ denotes the Gaussian distribution of real variable ${x}$ having mean of ${m}_{x}$ and variance of ${\lambda}_{x}$; $\mathcal{S}\backslash x$ denotes all variables in set $\mathcal{S}$ except $x$; $\mathbb{E}$ represents the expectation operator; $|\cdot|$ represents the modulus of a complex number; $\propto$ represents both sides of the equation are multiplicatively connected to a constant.

\section{System Model}
We consider a vehicular network with one RSU supporting $K$ vehicles. The RSU operates at mmWave band equipped a massive MIMO uniform linear array (ULA) which has $N_t$ transmit antennas and a separate array of $N_r$ receive antennas. This allows the RSU to receive the vehicle echoes for tracking while ensuring uninterrupted downlink transmission. Each vehicle is assumed to have an $M$-antenna ULA. Without loss of generality, we denote the range, the angle, and the speed of the $k$th vehicle relative to the RSU's array at the $n$th time instant are denoted by $d_{k,n}$, $\theta_{k,n}$, and $v_{k,n}$, respectively.

\subsection{Signal Model}
At the $n$th instant, the RSU sends a $K$-dimensional multi-beam direction-finding DFRC signal to the $K$ vehicles concurrently, denoted by $\tb{s}_n(t) = [s_{1,n}(t),...,s_{K,n}(t)]^{\rm T}$ with a complex signal $s_{k,n}(t)$ for vehicle $k$. The signal $\tb{s}_n(t)$ is transmitted over ${N}_t$ antennas of the RSU via transmit beamforming. In general, the beamforming matrix $\tb{F}_n$ is designed relying on the predicted angle. Assuming that we have a prediction of angle $\theta_{k,n}$ at time $n$, denoted by ${\theta}^{\rm pred, R}_{k,n}$, the beamforming vector for the $k$th vehicle is the $k$th column of $\tb{F}_n$, expressing as
\begin{align}\label{bf_vector}
\tb{f}_{k,n} = \sqrt{e_{k,n}}\tb{a}({\theta}^{\rm pred, R}_{k,n}),
\end{align}
where $e_{k,n}$ denotes the signal power, $\tb{a}({\theta}^{\rm pred, R}_{k,n})$ is the beam steering vector with the $i$th element being $a_i({\theta}^{\rm pred, R}_{k,n}) =e^{-j\pi (i-1)\cos {\theta}^{\rm pred, R}_{k,n}}$. The transmitted signal $\tilde{\tb{s}}_{k,n}(t)=\tb{f}_{k,n}s_{k,n}(t)$ is reflected by the $k$th vehicle and the received echo is denoted by $r_{k,n}(t)$. We assume that there is only negligible inter-beam interference and the RSU can identify the echoes from different vehicles. The reflected echo for the $k$th vehicle can then be expressed as
\begin{align}\label{rx_echo}
\tb{r}_{k,n}(t) = &\varsigma \beta_{k,n}e^{j2\pi \gamma_{k,n}t} \tb{b}(\theta_{k,n}) \tb{a}^{\rm H} (\theta_{k,n})\tilde{\tb{s}}_{k,n}(t-\tau_{k,n}) \nonumber\\&+ \tb{z}_{k,n}(t),
\end{align}
where $\varsigma=\sqrt{N_t N_r}$ denotes the multi-antenna array gain, $\beta_{k,n}$, $\gamma_{k,n}$, and $\tau_{k,n}$ denote the reflection coefficient, the Doppler, and the delay of the $k$th vehicle at time $n$, respectively, and $\tb{b}(\theta_{k,n})$ is the receive steering vector with the $i$th element being $b_i(\theta) =e^{-j\pi (i-1)\cos \theta}$. The term $\tb{z}_{k,n}(t)$ is assumed to be a complex additive white Gaussian noise with zero mean. Given the relative range of vehicle $k$ and the RSU, i.e., $d_{k,n}$, the reflection coefficient can be modeled as $\beta_{k,n}=\frac{\xi}{2{d_{k,n}}}$, where $\xi$ represents the complex radar cross-section (RCS)\cite{skolnik2001radar}.

By performing radar matched filtering on \eqref{rx_echo} with a delayed and Doppler-shifted version of $\tb{s}_{k,n}(t)$, we obtain the estimates of  delay $\tau_{k,n}$ and Doppler $\gamma_{k,n}$, which are related to range $d_{k,n}$ and speed $v_{k,n}$, respectively. And the measurement model for the motion parameters $d_{k,n}$ and $v_{k,n}$ are given by
\begin{align}\label{delay}
\tau_{k,n} &= \frac{2d_{k,n}}{c} + z_\tau,~\textrm{and}\\\label{doppler}
\gamma_{k,n} &= \frac{2v_{k,n} \cos \theta_{k,n} f_c}{c} + z_\gamma,
\end{align}
respectively, where $f_c$ and $c$ represent the carrier frequency and THE signal propagation speed, noise terms $z_\tau$ and $z_\gamma$ obey Gaussian distributions $\mathcal{N}(z_\tau;0,\sigma_\tau^2)$ and $\mathcal{N}(z_\gamma;0,\sigma_\gamma^2)$, respectively.

Having the estimates of delay and Doppler, we have the received signal samples ${\tb{y}}_{k,n}=[y_{k,n}^{[1]},...,y_{k,n}^{[N_r]}]^T$ for $\theta_{k,n}$ and $\beta_{k,n}$ based on the filtered signal, given by
\begin{align}\label{observation_y}
{\tb{y}}_{k,n} = \varsigma\beta_{k,n}\sqrt{e_{k,n}} \tb{b}({\theta}_{k,n})\tb{a}^\textrm{H}({\theta}_{k,n}) \tb{a}({\theta}^{\rm pred, R}_{k,n}) +{\tb{z}}_{k,n},
\end{align}
where the term $\tb{z}_{k,n} = [z_{k,n}^{[1]},...,z_{k,n}^{[N_r]}]^{\rm T}$ denotes the noise samples at different receive antennas. Without loss of generality, we model ${z}^{[i]}_{k,n}=z_y \sim\mathcal{N}(z_y;0,\sigma_y^2)$,~$\forall i,k,n$. Remark that after matched filtering, we achieve a signal-to-noise ratio (SNR) gain $G$, which is typically identical to the energy of signal $s_{k,n}(t)$.

\subsection{State Evolution Model}
 \begin{figure}[]
	\centering
	\includegraphics[width=.4\textwidth]{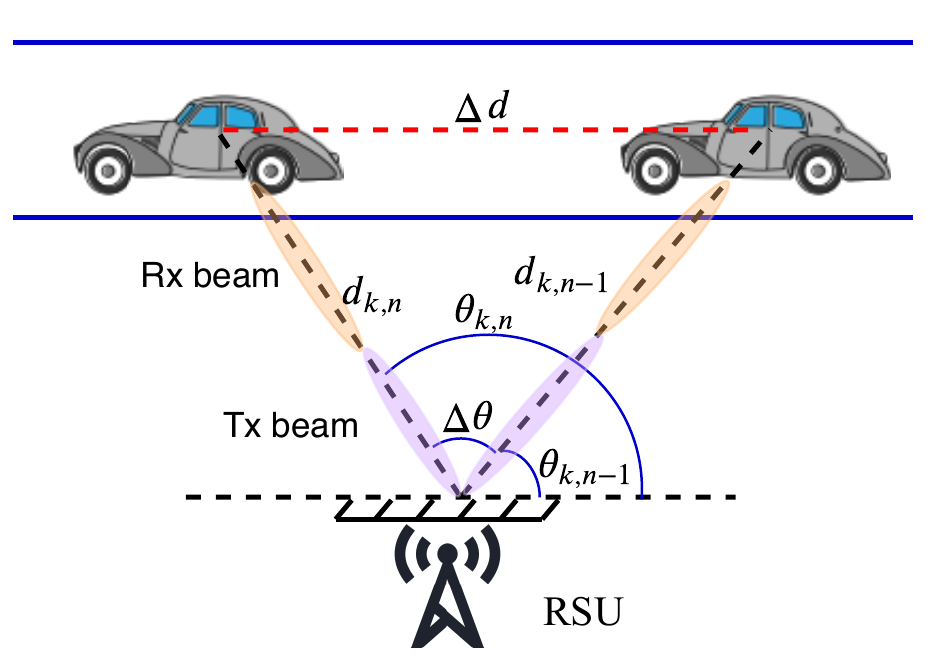}
	\caption{State evolution model of the considered vehicular network.}\label{model2}
	\centering
\vspace{-3mm}
\end{figure}

Based on the previous states and the moving patterns of the vehicles, we can determine the state evolution models for the vehicles, as shown in Fig. \ref{model2}. Relying on the geometric relationship of the motion parameters at time instant $n-1$ and $n$ as shown in Fig. \ref{model2}, we have the following kinematic equations as
\begin{align}\label{kinematic}
\left\{\begin{array}{l}
\sin (\theta_{k,n}-\theta_{k,n-1})d_{k,n} = v_{k,n-1} T \sin \theta_{k,n-1},\\
d_{k,n}^2 =  d_{k,n-1}^2 + (v_{k,n-1} T)^2 - 2d_{k,n-1}v_{k,n-1} T\cos \theta_{k,n-1}.
\end{array}
\right.
\end{align}
The above two equations show how the motion parameters of vehicle $k$ evolve with time. For brevity, we define $\Delta \theta = \theta_{k,n}-\theta_{k,n-1}$ and $\Delta d = v_{k,n-1} T$. Obviously, solving the above nonlinear equations to construct the evolution model is challenging. As a compromise approach, we propose to find a tractable approximation of \eqref{kinematic} by using the approximations $\Delta \theta_{k,n} \approx \sin\Delta \theta_{k,n}$, $d_{k,n}\approx d_{k,n-1}$, and $v_{k,n}\approx v_{k,n-1}$. Consequently, we summarize the state evolution models for $\theta_{k,n}$, range $d_{k,n}$, speed $v^{k,n}$, and coefficient $\beta_{k,n}$ in the following
\begin{align}\label{state_evolution_theta}
\theta_{k,n} &= \theta_{k,n-1}+\frac{v_{k,n-1} T \sin \theta_{k,n-1}}{d_{k,n-1}}+z_\theta,\\
\label{state_evolution_d}
d_{k,n} &= d_{k,n-1}-v_{k,n-1} T\cos \theta_{k,n-1}+ z_d,\\\label{state_evolution_v}
v_{k,n} &= v_{k,n-1}+z_v,\\\label{state_evolution_beta}
\beta_{k,n} &= \beta_{k,n-1}+\beta_{k,n-1}\frac{v_{k,n-1} T \cos \theta_{k,n-1}}{d_{k,n-1}}+z_\beta,
\end{align}
where the transition noise $z_\theta$, $z_d$, $z_v$, and $z_\beta$ obey zero mean Gaussian distributions $\mathcal{N}(z_\theta;0,\sigma_\theta^2)$, $\mathcal{N}(z_d;0,\sigma_d^2)$, $\mathcal{N}(z_v; 0,\sigma_v^2)$, and $\mathcal{CN}(z_\beta; 0,\sigma_\beta^2)$, respectively. Note that the state evolution of $d_{k,n}$, $\theta_{k,n}$, and $\beta_{k,n}$ also depend on other variables. For simplicity, we adopt the estimates at time instant $n-1$, i.e., $\hat{v}_{k,n-1}$, $\hat{\theta}_{k,n-1}$, and $\hat{d}_{k,n-1}$ to replace the corresponding terms in \eqref{state_evolution_theta}-\eqref{state_evolution_beta}, such that the evolution of the motion parameters only depends on their own previous states.

\subsection{Communication Model}
To receive the signal sent by the RSU, vehicle $k$ adopts a receive beamformer $\tb{w}_{k,n}$ and the received signal is formulated as
\begin{align}
g_{k,n}(t) = \bar{\varsigma}\alpha_{k,n}\tb{w}^H_{k,n} \tb{u}(\theta_{k,n})\tb{a}^{\rm H}(\theta_{k,n})\tilde{\tb{s}}_{k,n}(t) + {z}_g(t),
\end{align}
where $\bar{\varsigma}=\sqrt{N_t M}$ is the array gain between the RSU and the vehicle, $\alpha_{k,n}$ is the channel pathloss coefficient, ${z}_g(t)$ is the additive white Gaussian noise term, and $\tb{u}(\theta_{k,n})$ denotes the receive steering vector of vehicle $k$, which has a similar definition as $\tb{a}(\theta)$. The beamformer $\tb{w}_{k,n}$ is designed based on the predicted angle of vehicle $k$ relative to the RSU at time instant $n$, i.e., $\tb{w}_{k,n} = \tb{u}({\theta}^{\rm pred}_{k,n})$. Assuming that the original transmitted signal ${s}_{k,n}(t)$ from the RSU has unit power, then the SNR of the received signal is given by
\begin{align}\label{SNR_exp}
\textrm{SNR}_{k,n} = \frac{\left|\bar{\varsigma}\alpha_{k,n}\tb{w}^H_{k,n} \tb{u}(\theta_{k,n})\tb{a}^{\rm H}(\theta_{k,n})\tb{f}_{k,n}\right|^2}{N_0},
\end{align}
where $N_0$ is the power spectral density (PSD) of ${z}_g(t)$. Based on the SNR corresponding to the $k$th vehicle, the achievable sum-rate of all vehicles at time $n$ is expressed as $R_n = \sum_{k=1}^K (1+\textrm{SNR}_{k,n})$. It can be observed that when the angle is perfectly predicted, i.e., $\theta_{k,n} = {\theta}^{\rm pred, R}_{k,n} = {\theta}^{\rm pred}_{k,n}$, the received SNR is maximized.

For clarity, we define vectors ${\tb{y}}_k = [{\tb{y}}^{\rm T}_{k,1},...,{\tb{y}}^{\rm T}_{k,N}]^{\rm T}$, $\bm{\tau}_k=[\tau_{k,1},...,\tau_{k,N}]^{\rm T}$, and $\bm{\gamma}_k = [\gamma_{k,1},...,\gamma_{k,N}]^{\rm T}$ as the received signals, observed delays and Dopplers of vehicle $k$ until time instant $N$, respectively. Furthermore, the unknown parameters corresponding to vehicle $k$ can also be rewritten in vector form as $\bm{\theta}_k = [\theta_{k,1},...,\theta_{k,N}]^{\rm T}$, $\tb{d}_k=[d_{k,1},...,d_{k,N}]^{\rm T}$, $\tb{v}_k=[v_{k,1},...,v_{k,N}]^{\rm T}$, and $\bm{\beta}_k = [\beta_{k,1},...,\beta_{k,N}]^{\rm T}$, respectively. As the echo signals reflected by different vehicles can be identified unambiguously, in what follows, we will omit the vehicle index `$k$' for brevity. In the next section, we will formulate a factor graph model to infer the variables representing the motion parameters of vehicles.

\section{Factor Graph Model}
From the Bayesian perspective, we aim for inferring the variables from the observations through the maximum \emph{a posteriori} (MAP) estimator,
\begin{align}\label{MAP_est}
\{\hat{\tb{d}},\hat{\bm{\theta}},\hat{\tb{v}},\hat{\bm{\beta}}\} = \arg\max_{\tb{d},\bm{\theta},\tb{v},\bm{\beta}}p(\tb{d},\bm{\theta},\tb{v},\bm{\beta}|\tb{y},\bm{\tau}, \bm{\gamma}),
\end{align}
where $p(\tb{d},\bm{\theta},\tb{v},\bm{\beta}|\tb{y},\bm{\tau}, \bm{\gamma})$ denotes the joint \emph{a posteriori} distribution. Nevertheless, solving \eqref{MAP_est} involves a multi-dimensional search, leading to an exponentially increased complexity \cite{yuan2019iterative}. As a suboptimal solution, we will resort to the factor graph framework to obtain the marginals of unknown variables by leveraging the conditional independency between variables. According to Bayes Theorem, the joint distribution is rewritten as
$p(\tb{d},\bm{\theta},\tb{v},\bm{\beta}|\tb{y},\bm{\tau},\bm{\gamma},\bm{\beta}) = p(\tb{y},\bm{\tau}, \bm{\gamma}|\tb{d},\bm{\theta},\tb{v}) p(\tb{d},\bm{\theta},\tb{v},\bm{\beta})$,
where $p(\tb{y},\bm{\tau}, \bm{\gamma}|\tb{d},\bm{\theta},\tb{v},\bm{\beta})$ and $p(\tb{d},\bm{\theta},\tb{v},\bm{\beta})$ are the likelihood function and the joint \emph{a priori} distribution, respectively. Let us consider the \emph{a priori} distribution first. Based on the state transition function in \eqref{state_evolution_theta}-\eqref{state_evolution_beta}, the joint \emph{a priori} distribution can be factorized as
\begin{align}\label{prior}
p(\tb{d},\bm{\theta},\tb{v},\bm{\beta})=&p(d_{0})p(\theta_{0})p({v}_{0})p({\beta}_{0})\prod_{n=1}^N p(d_{n}|d_{n-1}) \nonumber\\&\cdot p(\theta_{n}|\theta_{n-1})p(v_{n}|v_{n-1})p(\beta_{n}|\beta_{n-1}),
\end{align}
Without loss of generality, we model the initial distributions of $d_{0},~\theta_{0}$, $v_{k,0}$, and $\beta_0$ as Gaussian distributions $p(d_{0})=\mathcal{N}(d_0;m_{{d}_{0}},\lambda_{d_{0}})$, $p(\theta_{0})=\mathcal{N}(\theta_0;m_{{\theta}_{0}},\lambda_{\theta_{0}})$, $p(v_{0})=\mathcal{N}(v_0;m_{{v}_{0}},\lambda_{v_{0}})$, and $p(\beta_{0})=\mathcal{N}(\beta_0;m_{{\beta}_{0}},\lambda_{\beta_{0}})$.

For the joint likelihood function, since the received signals, observed delays and Dopplers are irrelevant given the variables, we can express the joint likelihood function as
$p(\tb{y},\bm{\tau}, \bm{\gamma}|\tb{d},\bm{\theta},\tb{v},\bm{\beta}) = p(\tb{y}|\bm{\theta},\bm{\beta}) p(\bm{\tau}|\tb{d}) p(\bm{\gamma}|\bm{\theta},\tb{v})$.
Considering the independent Gaussian noise terms for different time instants, $p(\tb{y},\bm{\tau}, \bm{\gamma}|\tb{d},\bm{\theta},\tb{v},\bm{\beta})$ can be factorized as
\begin{align}\label{likelihood_factor}
p(\tb{y},\bm{\tau}, &\bm{\gamma}|\tb{d},\bm{\theta},\tb{v},\bm{\beta}) = \nonumber\\&\prod_{n=1}^N  \Big[ p(\gamma_{n}|\theta_{n},v_{n})p(\tau_{n}|d_{n})\prod_{l=1}^{N_r} p(y_{n}^{[l]}|\theta_{n},\beta_n)
\Big],
\end{align}
where $p(\tau_{n}|d_{n})$ and $p(\gamma_{n}|\theta_{n},v_{k,n})$ obey the Gaussian distributions $\mathcal{N}(r_n;\frac{2 d_n}{c},\sigma_\tau^2)$ and $\mathcal{N}(\gamma_n;\frac{2v_{n} \cos \theta_{n} f_c}{c},\sigma_\gamma^2)$. Based on the signal model, the received signal $y_{n}^{[l]}$ at the $l$th receive antenna consists of $N_t$ components, which makes the inference problem very difficult. Hence we introduce an auxiliary variable $\epsilon_{n}^{[q]}$ satisfying $\epsilon_{n}^{[q]}=e^{-j\pi q \cos\theta_{n}}$ and $y_{k,n}^{[l]}=\sum_{i=1}^{N_t} a_i({\hat{\theta}}_{n}^{0}) \epsilon_{n}^{[i-l]}+z_{y}$. Based on the auxiliary variables, $p(y_{n}^{[l]}|\theta_{n})$ is given by
\begin{align}\label{likelihood_ykn}
p(y_{n}^{[l]}|\theta_{n},\beta_n) \propto &\exp\left(\frac{|y_{n}^{[l]}-\beta_n\sqrt{p_n}\sum_{i=1}^{N_t} a_i({\hat{\theta}}_{n}^{0}) \epsilon_{n}^{[i-l]}|^2}{2\sigma_y^2}\right)\nonumber\\&\cdot\underbrace{\delta(\epsilon_{n}^{[i-l]}-e^{-j\pi (i-l) \cos\theta_{n}})}_{\kappa_{i-l}}.
\end{align}

\begin{figure}[!t]
	\centering
	\includegraphics[width=.4\textwidth]{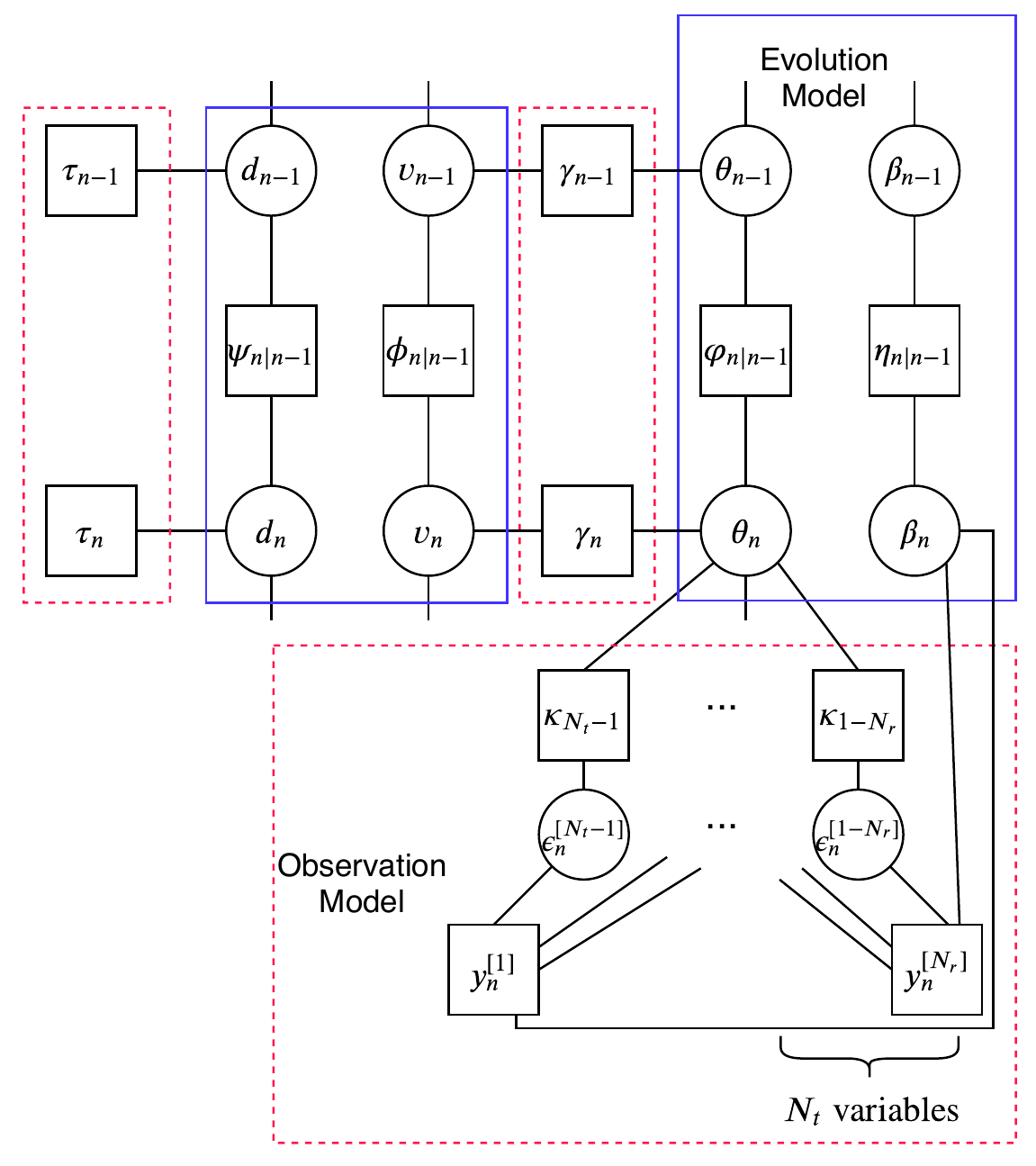}
	\caption{The factor graph representation of the considered problem. The shorthand notations $\psi_{n|n-1}$, $\phi_{n|n-1}$, $\varphi_{n|n-1}$, and $\eta_{n|n-1}$ denote the state transition probabilities $p(d_{n}|d_{n-1})$, $p(v_{n}|v_{n-1})$, $p(\theta_{n}|\theta_{n-1})$, and $p(\beta_n|\beta_{n-1})$, respectively. The factor vertices $\tau_{n}$, $\gamma_{n}$, and $y_n^{[i]}$ denote the likelihood functions corresponding to the observations $\tau_{n}$, $f_{n}$, and $y_n^{[i]}$.}\label{fig1}
	\centering
\vspace{-3mm}
\end{figure}

Following \eqref{prior}-\eqref{likelihood_ykn}, we have the factorization of the joint \emph{a posteriori} distribution and can represent it by a factor graph, as shown in Fig. \ref{fig1}.
Having the factor graph, message passing algorithm can be implemented to efficiently compute the ``beliefs'' (approximate marginals) of unknown variables, which will be elaborated in the following section.

\section{The Proposed Message Passing Approach}
There are two kinds of messages, i.e., the message from the factor vertex to the variable vertex and vice versa. We use $\rmsg{\mu}_{f}(x)$ to denote the message from the factor vertex $f$ to the variable vertex $x$ and $\lmsg{\nu}_{f}(x)$ to denote the message from $x$ to factor $f$. The message updating rules are defined in \cite{kschischang2001factor} and not given here for space limitation.

\subsection{Vehicle State Prediction}
We commence our discussions with the messages in the state evolution part. Provided that the belief of $v_{n-1}$ has been obtained in Gaussian form as $b(v_{n-1})=\mathcal{N}(v_{n-1}; m_{v_{n-1}},\lambda_{{v}_{n-1}})$, the message $\rmsg{\mu}_{\phi_{n|n-1}}(v_n)$ is given by
\begin{align}
&\rmsg{\mu}_{\phi_{n|n-1}}(v_n)\propto \exp\left(-\frac{(v_n-m_{v_{n-1}})^2}{2(\sigma_v^2+\lambda_{{v}_{n-1}})}\right).
\end{align}
It can be observed that the above message subjects to Gaussian distribution, which is characterized by the mean ${m}_{\phi_{n|n-1}\to v_n}=m_{v_{n-1}}$ and variance ${\lambda}_{\phi_{n|n-1}\to v_n}=\sigma_v^2+\lambda_{{v}_{n-1}}$. Therefore, we use the corresponding mean and variance to simplify the message derivations.

In a similar way, we can derive the messages $\rmsg{\mu}_{\psi_{n|n-1}}(d_n)$, $\rmsg{\mu}_{\eta_{n|n-1}}(\beta_n)$, and $\rmsg{\mu}_{\varphi_{n|n-1}}(\theta_n)$ related to the vehicle state predication, expressing as
\begin{align}\label{eq44}
\left\{
\begin{array}{l}
{m}_{\psi_{n|n-1} \to d_n}=m_{d_{n-1}}-\hat{v}_{n-1} T\cos \hat{\theta}_{n-1},\\
{\lambda}_{\psi_{n|n-1} \to d_n}=\sigma_d^2+\lambda_{d_{n-1}},\\
{m}_{\varphi_{n|n-1}\to\theta_n}=m_{\theta_{n-1}}+\frac{\hat{v}_{n-1} T\sin \hat{\theta}_{n-1}}{\hat{d}_{n-1}},\\
{\lambda}_{\varphi_{n|n-1}\to\theta_n}=\sigma_\theta^2+\lambda_{\theta_{n-1}},\\
{m}_{\eta_{n|n-1}\to\beta_n}=\rho_{n-1}\rho_{n-1} m_{\beta_{n-1}},\\
{\lambda}_{\eta_{n|n-1}\to\beta_n}=\sigma_\beta^2+\rho_{n-1}^2 \lambda_{\beta_{n-1}}.
\end{array}\right.
\end{align}
It can be observed that the means and variances in \eqref{eq44} are updated based on the marginal mean and variance in the previous time instant and the state evolution model.
Based on the Gaussian form message $\rmsg{\mu}_{\varphi_{n|n-1}}(\theta_n)$, we have the predicted angle of ${\theta}^{\rm pred,R}_{n}$ at the $n$th epoch as
\begin{align}\label{predRSU}
{\theta}^{\rm pred,R}_{n}=\arg\max_\theta \rmsg{\mu}_{\varphi_{n|n-1}}(\theta_n) = {m}_{\varphi_{n|n-1}\to\theta_n},
\end{align}
which is used for designing the beamformer at the RSU.

\subsection{Vehicle State Tracking}
In the following, we will discuss the message calculations related to the observation model.
\subsubsection{Messages related to $\tau_n$}
The message $\rmsg{\mu}_{\tau_n} (d_n)$ is identical to the likelihood function $p(\tau_n|d_n)$ since $\tau_n$ depends solely on $d_n$. After straightforward manipulations, we write $\rmsg{\mu}_{\tau_n} (d_n)$ as
\begin{align}
\rmsg{\mu}_{\tau_n} (d_n) \propto \mathcal{N}\left(d_n;\frac{c\tau_n}{2},\frac{\sigma_\tau^2\,c^2}{4}\right).
\end{align}
Then the belief of $d_n$ at time instant $n$ can be obtained as $b(d_n) \propto \mathcal{N}\left(d_n;m_{d_n},\lambda_{d_n}\right)$
with the mean and variance being
\begin{align}
m_{d_n} &= \lambda_{d_n}\left(\frac{2c\tau}{\sigma_\tau^2\,c^2}+\frac{m_{d_{n-1}}-\hat{v}_{n-1} T\cos \hat{\theta}_{n-1}}{\sigma_d^2+\lambda_{d_{n-1}}}\right),\\
\lambda_{d_n} &= \left(\frac{4}{\sigma_\tau^2\,c^2}+\frac{1}{\sigma_d^2+\lambda_{d_{n-1}}}\right)^{-1}.
\end{align}
Since $b(d_n)$ is a Gaussian distribution, the estimate of range $d_n$ is $\hat{d}_{n} = m_{d_n}$. The estimate $\hat{d}_{n}$ is used for modeling the state evolution function. Also, the obtained belief is passed to factor vertex $\psi_{n+1|n}$ for calculating $\rmsg{\mu}_{\psi_{n|n-1}}(d_n)$.

\subsubsection{Messages related to $\gamma_n$}
\begin{figure}[]
	\centering
	\includegraphics[width=.45\textwidth]{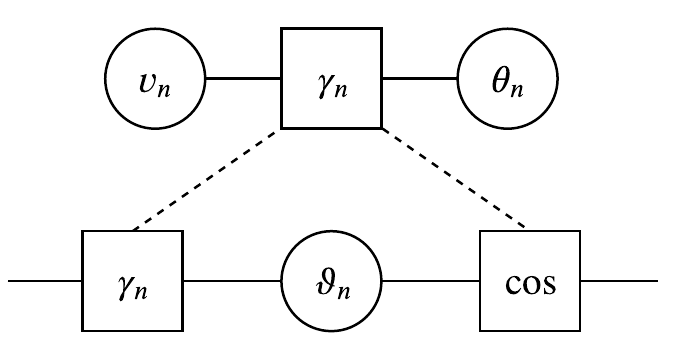}
	\caption{Reconstruction of the factor vertex $\gamma_n$.}\label{fig2}
	\centering
\vspace{-3mm}
\end{figure}

Note that \eqref{doppler} involves a nonlinear cosine function, calculating the message $\rmsg{\mu}_{\gamma_n}(v_{n-1})$ can not provide a closed-form expression. To tackle this problem, we reconstruct the factor node $\tau_n$ by introducing a factor vertex representing the cosine function and a variable vertex denoting the cosine of an angle, as illustrated in Fig. \ref{fig2}.
To obtain a Gaussian closed-form message $\rmsg{\mu}_{\gamma_n}(v_{n})$, we introduce the mean field (MF) message passing \cite{winn2005variational} such that the mean and variance of $\rmsg{\mu}_{\gamma_n}(v_{n})$ are
\begin{align}
m_{\gamma_n\to v_n} &= \frac{\gamma_n m_{\vartheta_{n}\to \gamma_n}}{C_1(\lambda_{\vartheta_{n}\to \gamma_n}+m_{\vartheta_{n}\to \gamma_n}^2)},~C_1 = \frac{2 f_c}{c}\\
\lambda_{\gamma_n \to v_n} &= \frac{\sigma_\gamma^2}{C_1^2(\lambda_{\vartheta_{n}\to \gamma_n}+m_{\vartheta_{n}\to \gamma_n}^2)},
\end{align}
respectively. The belief of $v_n$ can be obtained with message $\rmsg{\mu}_{\gamma_n}(v_{n})$ and $\rmsg{\mu}_{\phi_{n|n-1}}(v_n)$ and the estimate $\hat{v}_n$ is used for predicating the parameters in the $(n+1)$th epoch.

For message $\rmsg{\mu}_{\cos}(\theta_n)$, the function turns to be an inverse cosine function. To overcome this nonlinear issue, we employ the second order Taylor expansion concerning the inverse cosine function as $\arccos\vartheta\approx \pi/2-\vartheta-\vartheta^3/6$. Then based on the obtained parameter $m_{\gamma_n\to \vartheta_n}$ and $\lambda_{\gamma_n \to \vartheta_n}$, we derive the Gaussian message $\rmsg{\mu}_{\gamma_n}(\theta_n)$ with mean
$m_{\cos \to \theta_n} = \frac{\pi}{2}-m_{\gamma_n\to \vartheta_n}-\frac{m_{\gamma_n\to \vartheta_n}^3+3m_{\gamma_n\to \vartheta_n}\lambda_{\gamma_n \to \vartheta_n}}{6}$ and variance $\lambda_{\cos \to \theta_n} = \mathbb{E}[\arccos^2\vartheta_n]- m_{\cos \to \theta_n}^2$.

\subsubsection{Messages related to $y_{n}^{[l]}$}
Next, we derive the messages related to the observations $y_{n}^{[l]}$. Assuming that all messages from $\epsilon_{n}^{[q]}$ to $y_{n}^{[l]}$ are known with Gaussian distributions, it is readily to obtain the message $\rmsg{\mu}_{y_{n}^{[l]}}(\beta_n)$ using MF rules with mean $m_{y_{n}^{[l]} \to \beta_n}$ and variance $\lambda_{y_{n}^{[l]} \to \beta_n}$, respectively. The mean $m_{y_{n}^{[l]} \to \beta_n}$ and $\lambda_{y_{n}^{[l]} \to \beta_n}$ represents the information of the observation at the $l$th receive antenna contributed to the variable $\beta_n$. Having $\rmsg{\mu}_{y_{n}^{[l]}}(\beta_n)$ in hand, the belief of $\beta_n$ can be obtained  with mean and variance
\begin{align}
m_{\beta_n} &= v_{\beta_n}\left(\frac{m_{\eta_{n|n-1}\to\beta_n}}{\lambda_{\eta_{n|n-1}\to\beta_n}}+\sum_{l=1}^{N_r}\frac{m_{y_{n}^{[l]} \to \beta_n}}{\lambda_{y_{n}^{[l]} \to \beta_n}}\right),\\
\lambda_{\beta_n} &= \left(\frac{1}{\lambda_{\eta_{n|n-1}\to\beta_n}}+\sum_{l=1}^{N_r}\frac{1}{\lambda_{y_{n}^{[l]} \to \beta_n}}\right)^{-1}.
\end{align}

Finally, we aim for computing the messages related to the function $\kappa_q$, which involves the nonlinear function $e^{-jq\cos\theta_n}$. Note that this part of factor graph has cycles, we have to implement the message passing algorithms for a few iterations\cite{ihler2005loopy}. As above, the mean and variance of $\tilde{\theta}_n=\cos\theta_n$ is denoted by ${m}_{\tilde{\theta}_n\to \kappa_q}$ and ${\lambda}_{\tilde{\theta}_n\to \kappa_q}$. Employing the transformations of trigonometric functions and after some manipulations, we have the mean and variance for message $\rmsg{\mu}_{\kappa_q}(\epsilon_n^{[q]})$, formulating as
\begin{align}\label{m_kappa}
m_{\kappa_q \to \epsilon_n^{[q]}} &=e^{-{q^2 {\lambda}_{\tilde{\theta}_n\to \kappa_q}}/{2}}e^{-j q {m}_{\tilde{\theta}_n\to \kappa_q}},\\\label{v_kappa}
\lambda_{\kappa_q \to \epsilon_n^{[q]}} &= 1-e^{-q^2 {\lambda}_{\tilde{\theta}_n\to \kappa_q}}.
\end{align}
Then, to calculate the message $\rmsg{\mu}_{\kappa_q}(\theta_n)$ in Gaussian closed-form, we reconstruct the function node $\kappa_n$ by and separate $e^{-jq\tilde{\theta}_n}$ as $\cos(q\tilde{\theta}_n)+j\sin(q\tilde{\theta}_n)$. The real and imaginary parts of the mean as well as the variance of $\lmsg{\nu}_{\kappa_{q}}(\epsilon_n^{[q]})$ are passed to the inverse cosine (``acos'') and inverse sine (``asin'') functions, respectively. Based on Taylor expansion, the means and variances of the messages $\rmsg{\mu}_{\text{asin}}(\tilde{\theta}_n)$ and $\rmsg{\mu}_{\text{acos}}(\tilde{\theta}_n)$ can be obtained similar to $\rmsg{\mu}_{\gamma_n}(\theta_n)$. Consequently, we can calculate all messages form $\kappa_q$ to $\theta_n$ and arrive at the belief of $\theta_n$,
\begin{align}
b(\theta_n) = \rmsg{\mu}_{\gamma_n}(\theta_n)\rmsg{\mu}_{\varphi_{n|n-1}}(\theta_n)\prod_q \rmsg{\mu}_{\kappa_q}(\theta_n),
\end{align}
which is used for beam prediction at $(n+1)$th time instant.

In the above, we have solved the beam prediction and beam tracking problems based on the factor graph framework. With the help of the determined angular parameters, the RSU and the vehicle can maintain a reliable link for data transmission.

\section{Simulation Results}
\begin{figure}[!t]
    \centering
    \includegraphics[width=1\columnwidth]{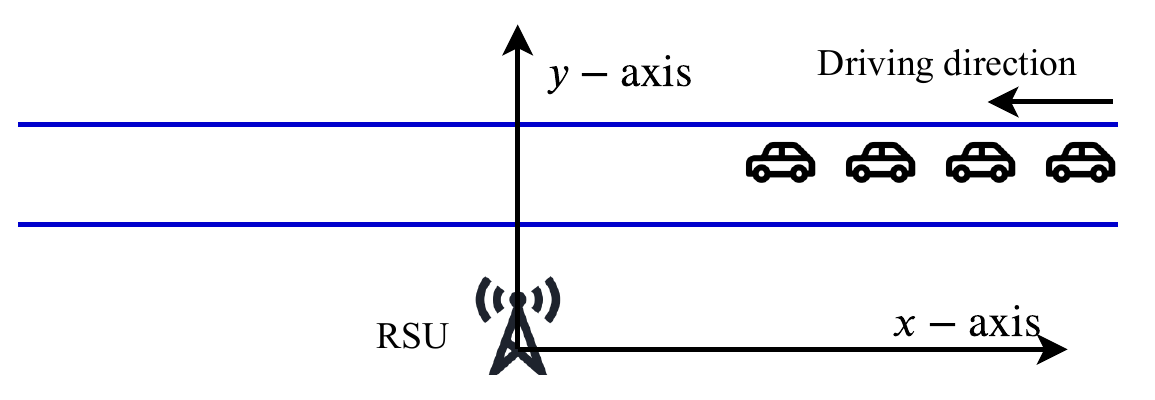}
    \caption{{The considered vehicular network for simulations.}}
    \label{simu_model}
    \vspace{-3mm}
\end{figure}
Let us consider a network with 4 vehicles moving on the road, as shown in Fig. \ref{simu_model}. Without loss of generality, the coordinate of the RSU is set as $[0,0]^{\rm T}$ and the initial positions of vehicles are $[100, 20]^{\rm T}$, $[90, 20]^{\rm T}$, $[80, 20]^{\rm T}$, and $[70, 20]^{\rm T}$, respectively. The RCS $\xi$ is set to $10+10j$, which is used for calculating the reflection coefficient $\beta_{k,0}$. The speeds of four vehicles at time instant $0$ are randomly generated from the uniform distribution $[10,20]$ m/s. The RSU and the vehicles are operating at a carrier frequency of $f_c=30$ GHz. The time slot duration is $T=0.02~s$ and the signal propagation speed is approximated as $c=3\times10^{8}$ m/s. For brevity, we set both the radar noise variance $\sigma_y^2$ and the noise PSD for communication $N_0$ to 1. For the observed delays and Dopplers at RSU, we use the standard deviations of $\sigma_\tau = 0.67~\mu$s and $\sigma_\gamma = 2$ kHz for all vehicles at different time slots. The state transition noises are set with standard deviations of $\sigma_d = 0.2$ m, $\sigma_v = 0.5$ m/s, $\sigma_\beta = 1$, and $\sigma_\theta = 0.02^{\circ}$. All results are averaged from 1,000 independent Monte Carlo simulations.


\begin{figure}[!t]
    \centering
    \includegraphics[width=0.9\columnwidth]{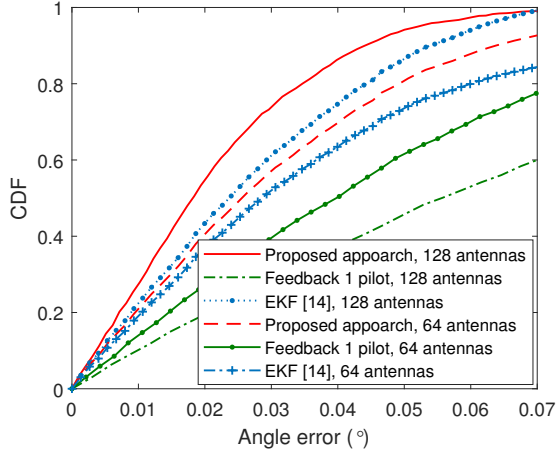}
    \caption{CDF of the angle estimation error.}\label{CDF_angle}
    \vspace{-3mm}
\end{figure}
\begin{figure}[!t]
    \centering
    \includegraphics[width=0.9\columnwidth]{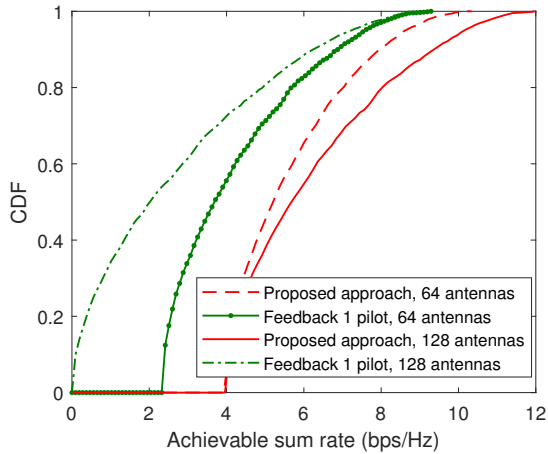}
    \caption{CDF of the communication achievable rate.}\label{CDF_rate}
    \vspace{-3mm}
\end{figure}
We compare the performance of the proposed approach and the classic feedback-based algorithm. Note that in the feedback-based scheme, the pilot are contained in the downlink communication signal. In contrast to the DFRC signal that the whole block can be used as the pilots, the feedback scheme employ only 1 pilot, leading to a much smaller SNR gain after matched filtering. For simplicity, we equivalently multiply the noise variance $\sigma_y^2$ by a constant for the feedback-based scheme.
{Fig. \ref{CDF_angle} evaluates the angle estimation result using the proposed approach, the feedback scheme, and the EKF method in \cite{liu2020radar}. We illustrate the CDF versus the angle estimation error at the last time instant for 1,000 trails. Two cases with 64 and 128 antennas are illustrated. We see that for both cases, the feedback scheme suffers from a remarkable performance loss due to limited matched-filtering gain. In contrast to the proposed approach, increasing the number of antennas for the feedback scheme will lead to performance degradation. This observation can be explained by the fact that a higher number of antennas provides a narrower beam, in which case using only 1 pilot is not sufficient to track the variation of the angular parameter. Moreover, the proposed algorithm outperforms the EKF method since EKF employs only the first-order Taylor expansion and neglect the higher-order information.} Since the estimated angles are used for beamforming design, the tracking error of angles will result in the misalignment of the beams. As a consequence, the received SNR is reduced, leading to a lower achievable rate. In Fig. \ref{CDF_rate}, we illustrate the CDF of the communication achievable rate of all time instants based on the proposed and the feedback-based methods at a SNR of 10 dB. For the proposed scheme, the achievable rate $R$ at different time instants are higher than 4 bps/Hz. While the achievable rate for the feedback-based scheme is much lower. This validates our discussions above that the large angle estimation error in the feedback-based approach degrades the achievable rate. Furthermore, the rate degradation becomes more significant for the feedback-based scheme in the case with 128 antennas, where the angle variation cannot be accurately tracked due to the narrow beamwidth. Figs. \ref{CDF_angle} and \ref{CDF_rate} show the superiority of employing DFRC signaling for reliable communication in vehicular networks.

\section{Conclusions}
In this paper, we proposed a novel DFRC based predictive beamforming scheme for vehicular networks, which has the advantages of lower signaling overhead and better performance than the conventional communication-only feedback-based scheme. We commence from a Bayesian perspective and construct the joint \emph{a posteriori} distribution based on the echo signals received at the RSU and the state transition models of the vehicles. Then the message passing algorithm is utilized to estimate the unknown variables. With appropriate approximations, the messages on factor graph were determined in closed-form, providing a low complexity solution for the considered beam tracking problem. Simulation results demonstrate the effectiveness and superiority of the proposed approach compared to the feedback-based scheme.

\bibliographystyle{IEEEtran}
\bibliography{v2x}
\end{document}